\documentclass[12pt]{article}

\usepackage{pstricks}     
\usepackage{graphicx}     
\usepackage{axodraw}   
\usepackage[latin2]{inputenc}   
\usepackage{epic}     
\usepackage{latexsym}   
\usepackage{epsf}      
\usepackage{epsfig}    
\usepackage{amssymb,amsmath} 

\newcommand{\sfrac}[2]{{\textstyle\frac{#1}{#2}}}
\newcommand{\binomial}[2]{ {\textstyle {#1 \choose #2 }}}

\newcommand{\cref}[1]{Chapter \ref{c.#1}}

\def\beq{\begin{equation}} 
\def\eeq{\end{equation}} 
\def\bea{\begin{eqnarray}}  
\def\eea{\end{eqnarray}}  
\def\ba{\begin{array}}  
\def\ea{\end{array}}   
\def\bi{\begin{itemize}}  
\def\ei{\end{itemize}}  
\def\be{\begin{enumerate}}  
\def\ee{\end{enumerate}}  
\def\bc{\begin{center}}
\def\ec{\end{center}}  

\def\pa{\partial}  
\def\hc{{\rm h.c.}} 
\def\bn{{(n)}}

\def\bmn{{(-n)}}

\def\bz{{(0)}}

\def\cl{{\mathcal L}}

\def\t{\tilde}

\def\nn{\nonumber \\}

\newsavebox{\moose} 
\sbox{\moose}{%
\begin{picture}(0,0)
  \thicklines
  \put(-60,0){\circle{35}}
  \put(60,0){\circle{35}}
  \ArrowLine(-40,0)(40,0)
\end{picture}} 

\newsavebox{\site}
\sbox{\site}{%
\begin{picture}(0,0)
  \thicklines
  \put(0,0){\circle{40}}
 \ArrowLine(-30,0)(-10,0)
  \ArrowLine(10,0)(30,0)
\end{picture}}

\newsavebox{\dsite}
\sbox{\dsite}{%
\begin{picture}(0,0)
  \thicklines
  \put(-60,0){\circle{40}}
  \put(60,0){\circle{40}}
  \ArrowLine(-40,0)(40,0)
\end{picture}}

\begin{document}

\begin{titlepage}

\begin{flushright}
{\tt hep-ph/0310201} \\
Saclay t03/152\\
IFT-03/29\\
\end{flushright}

\vskip.5cm

\begin{center}
{\huge \bf Loop Corrections in Higher}
\vskip.1cm
{\huge \bf  Dimensions via Deconstruction}
\end{center}
\vskip0.2cm

\begin{center}
{\bf
{Adam  Falkowski}$^{a}$,
{Christophe Grojean}$^{b,c}$
{\rm and}
Stefan  Pokorski$^{a}$}
\end{center}
\vskip 8pt

\begin{center}
$^{a}$ {\it Institute of Theoretical Physics, Warsaw University
Ho\.za 69, 00-681 Warsaw, Poland} \\
\vspace*{0.1cm}
$^{b}$ {\it Service de Physique Th\'eorique,
CEA Saclay, F91191 Gif--sur--Yvette, France} \\
\vspace*{0.1cm}
\hspace{-.7cm}
$^{c}$ {\it Michigan Center for Theoretical Physics, 
University of Michigan, Ann Arbor MI 48109, USA}\\
\vspace*{0.3cm}
{\tt afalkows@fuw.edu.pl,  grojean@spht.saclay.cea.fr, pokorski@fuw.edu.pl}
\end{center}

\vglue 2.5truecm

\begin{abstract}
\vskip 3pt
\noindent 
We calculate the one-loop corrections to the Kaluza-Klein gauge boson excitations in the deconstructed version of the 5D QED.  Deconstruction provides  a renormalizable UV completion of the 5D theory
that enables to control the cut-off dependence of  5D theories and study a possible influence of UV physics on IR observables.  In particular we  calculate the cut-off-dependent non-leading corrections that  may be phenomenologically relevant for collider physics. We also discuss the structure of  the operators that are relevant for the quantum corrections to the gauge boson masses in 5D and in deconstruction.    
\end{abstract}


\end{titlepage}

\newpage

\newpage 

\setcounter{page}{1} \pagestyle{plain}

\section{Introduction}
 
 In the past few years, High Energy Physics ventured to explore phenomenological  aspects of space-times involving more than four dimensions. From the hierarchy to the flavor problem, from supersymmetry to electroweak symmetry breaking, from  proton stability to the number of the Standard Model generations, from  Dark Matter abundance to neutrino oscillations, many puzzles that jeopardize our 4D understanding of Quantum Field Theory could find a  solution when extra dimensions are involved. So one is naturally led to wonder what is so special about extra dimensions? The notion of locality/sequestering is definitively an essential tool in suppressing any dangerous radiative operator. It was then  realized~\cite{arcoge} that locality in physical extra dimension can be advantageously mimicked by locality in theory space along which 4D gauge symmetry is multi-replicated. At tree-level, by a matching in the IR  of the mass spectra and the interaction patterns, a precise
correspondence has been established between higher dimensional theories and
4D deconstructed theories. This correspondence is believed to hold all the way long from the perturbative to the non-perturbative regime~\cite{NP}. 

Higher dimensional gauge  theories are non-renormalizable and valid only below certain physical cut-off scale $\Lambda$. Calculating quantum corrections in such theories requires a careful choice of a regularization scheme as, in general, there is a clash between the gauge invariance and the need for a cut-off~\cite{5Dreg}. The question of regularization  arises even for  those radiative corrections that are expected to be UV finite ({\it i.e.} dominated by IR physics).
Deconstruction can serve as a renormalizable UV completion of higher dimensional gauge theories and, within such scheme, calculation of quantum correction is totally unambiguous. Moreover, in deconstructed theory, radiative corrections include the effects due to a finite cut-off $\Lambda$. Although they are specific for this particular UV completion, they illustrate how the predictions of higher dimensional theories can be disturbed by UV physics. 
 
Recently, one-loop corrections to the masses of the gauge boson excitations have been calculated \cite{chmasc,geirqu,kupu}. In the present paper we calculate analogous corrections in the renormalizable deconstruction set-up and compare the results. We will restrict ourselves to 5D QED compactified on a circle (see~\cite{hile} for the corresponding setup),
the group theory factors associated to the non-abelian nature of the interactions being identical in the 5D and the 4D computations anyway. In this simple case, it was shown in Refs.~\cite{chmasc,geirqu,kupu} that
the interactions with a single 5D fermion of  electric charge $e_5$ shift
all the masses of the 4D KK gauge bosons  by an amount
\begin{equation}
        \label{e.dmamu}
\delta m^2_n = - \frac{ \zeta (3)\, e_0^2}{4 \pi^4 R^2} ,
\end{equation}
where $e_0=e_5/\sqrt{2\pi R}$ is the 4D gauge coupling and $R$ is the radius of the compact fifth dimension
(the 4D massless photon remains of course massless by gauge invariance).
Meanwhile, the massless scalar field corresponding to the component of the 5D gauge field along the  compact dimension acquires a mass given by \cite{ho}
\begin{equation}
        \label{e.dma5}
\delta m^2 = - \frac{ 3 \zeta (3)\, e_0^2}{4 \pi^4 R^2}. 
\end{equation}

The phenomenological relevance of one-loop corrections to the 4D gauge boson masses in 5D gauge theories has been stressed in Ref.~\cite{chmasc} where it was noticed that, due to the degeneracy of KK spectrum at tree-level, decay channels are controlled
by radiative corrections, thus a slight modification in the modification, in particular from UV physics,  
can affect collider signals~\cite{chmasc} as well as the abundance of Dark Matter~\cite{KKDM}. 
Thus the importance
of our computation in the deconstruction regularization where
we have a full control on the UV physics.
Let us also mention that in models where the Higgs boson is identified as a component of a gauge boson in higher dimensions~\cite{HiggsA5}, the radiative corrections we are interested in
ultimately control the electroweak symmetry breaking and determine the Higgs mass.
Finally, computing the radiative corrections to gauge bosons masses in 4D deconstructed theories is also important for the following reason: in~Ref.~\cite{csergr}, it was shown that the spectrum
of a product of $\mathcal{N}=1$ supersymmetric $SU(N)$ gauge theories broken to the diagonal
$SU(N)$ exhibit a $\mathcal{N}=2$ supersymmetry. Even though this extended supersymmetry
seems accidental from the 4D point of view, it is actually dictated by the underlying 5D Lorentz invariance of the corresponding higher dimensional theory. Our computation can be extended to show
that  the $\mathcal{N}=2$ supersymmetry indeed survives at one-loop.
 
\section{Framework}

\subsection{Tree-level matching between the 5D and 4D theories}

As outlined in the introduction, we restrict ourselves to the  case of 5D QED and a massless Dirac fermion of electric charge $e$, the fifth dimension being compactified on a circle of radius $R$. The deconstructed setup (see also~\cite{hile})
corresponds to a product of  $N$ copies of $U(1)$ gauge group\footnote{For definiteness we take $N$ to be odd.} linked together by $N$ scalar fields $\Phi_p$ of charge $(e,-e)$ under $U(1)_p\times U(1)_{p+1}$ (the site indices being periodically identified as $N+p \sim p$).  Once the link fields acquire a VEV, $\langle \Phi_p \rangle =v/\sqrt{2}$, the product gauge group is broken to the diagonal
$U(1)$ and the gauge boson spectrum is made of a massless photon:
\begin{equation}
         \label{eq:m0gauge}
A_\mu^\bz = \frac{1}{\sqrt{N}} \sum_{p=1}^N A_{\mu,p},
\end{equation}
and a tower of massive excitations doubly degenerated in mass ($n=1,\ldots, (N-1)/2$)
\begin{equation}
        \label{eq:mkgauge}
A_\mu^\bn = \sqrt{\frac{2}{N}} \sum_{p=1}^N \cos \frac{2n(p-1)\pi}{N} A_{\mu,p} \ \mathrm{and}\
A_\mu^\bmn = \sqrt{\frac{2}{N}} \sum_{p=1}^N \sin \frac{2n(p-1)\pi}{N}  A_{\mu,p}
\end{equation}
with mass
\begin{equation}
        \label{eq:masses}
m_{\pm n} = 2\, e v  \sin \frac{n \pi}{N} \, .
\end{equation}
The shift symmetry of the setup, {\it i.e.}, the fact that the electric charges and VEVs do not depend on the site index, corresponds to the  translational symmetry of the fifth dimension compactified on
a circle.

The deconstruction set-up can be thought of as a discretization of the fifth dimension
at points $y_p=2p \pi R/N$, $p=1\ldots N$, the 5D gauge field
being matched to the 4D degrees of freedom in the following way.
The 4D components of the gauge field at the point $y_p$, $A_\mu(x_\nu,y_p)$, 
are identified with the 4D gauge field at the site $p$, $A_{\mu,p}(x_\nu)$.
The component along the extra dimension of the 5D gauge field, $A_5(x_\nu,y_p)$, is matched to the link field $\Phi_p(x_\nu)$, as it can be seen in the broken phase of the deconstruction theory.
Indeed let us split the link fields as $\Phi_p = {1 \over \sqrt 2}( v{\bf I} +\Sigma_p + i G_p)$. For a number of sites large enough, a gauge invariant renormalizable scalar potential can depend on the link fields only in the combination $\Phi_p^*\Phi_p$. In consequence, the scalar sector of the theory possesses an additional  $U(1)^N$ global symmetry (acting as $\Phi_p \to e^{ i \alpha_p}\Phi_p$), which is completely broken  when  the links acquire VEVs.
This global symmetry pattern  results in the presence of $N$ massless Goldstone bosons, $N-1$ of which actually being eaten by the massive gauge bosons. The remaining physical Goldstone boson, identified as $G_\bz = (G_1 + \dots + G_N)/\sqrt{N}$, is precisely what matches the zero mode of $A_5$.  Finally, the real parts of the link fields, $\Sigma_p$, can acquire a mass of the order of the deconstruction scale and thus they do not match any degrees of freedom of the  5D theory below its cutoff $\Lambda$.

To reproduce the fermionic KK modes, we need to introduce
$N$ pairs of chiral fermions $(\psi_{p},\chi_{p})_{p=1\ldots N}$ of charge $(e,e)$ under $U(1)_p$.
After the breaking to the diagonal $U(1)$, 
the correct KK spectrum is recovered in the large $N$ limit 
at the condition to correctly fine-tune the Yukawa couplings of the fermions as follows~\cite{csergr}:
\begin{eqnarray}
\mathcal{L}  & = & 
\sum_{p=1}^N 
\left (
i {\bar \psi}_{p} \sigma^\mu D_{\mu,p}  \psi_{p} +
i {\bar \chi}_{p} \sigma^\mu  D_{\mu,p} \chi_{p}
 + \sqrt{2}\, e\, \Phi_p {\bar \chi}_{p} \psi_{p+1} - e\, v\, {\bar \chi}_{p} \psi_{p}  
 + \mathrm{h.c.} 
 \right)  , 
\end{eqnarray}
where $D_{\mu,p}$ stands for the covariant derivative for the $U(1)_p$ gauge group,
$D_{\mu,p} =\partial_\mu + i e A_{\mu,p}$. After symmetry breaking down to the diagonal 
$U(1)$, the fermionic spectrum is made of one massless Dirac fermion and
a tower of massive Dirac fermions with the same mass as the gauge boson ones (see~\cite{fgp} for details about the mode decomposition). Note that due 
to the normalization factor appearing in the massless photon~(\ref{eq:m0gauge}), all these fermions
carry a charge $e_0=e/\sqrt{N}$ under the unbroken $U(1)$ gauge group.

The comparison of the spectrum and the interactions in both the compactified 5D theory and
the deconstructed 4D theory leads to the following identification of the parameters~\cite{arcoge}
\begin{equation}
        \label{e.matching}
e_0=\frac{e_5}{\sqrt{2\pi R}} = \frac{e}{\sqrt{N}}  
\ \ \mathrm{and}\ \
\frac{1}{R} = \frac{2\pi}{N}\, ev.
\end{equation}
The cutoff scale, $\Lambda$, of the 5D theory is also related to the 4D parameters by $\Lambda = e v$.

\subsection{Renormalization set-up}

At the quantum level, the 4D deconstructed theory constitutes a UV completion of the 5D gauge theory, and the framework can be arranged to be renormalizable.   Therefore, at an arbitrary level of perturbation theory,  all observables are unambiguously determined up to the freedom of adjusting a finite number of counterterms. Note that the form of the counterterms  is additionally constrained by  the discrete shift symmetry inherited from the 5D translational invariance.

The bare and renormalized quantities  are related to one another as follows:
\begin{equation} 
        \label{e.dct}
A_{\mu,p}^B = Z_A^{1/2} A_{\mu,p}, \
\Phi_p^B = Z_\Phi^{1/2} \Phi_p,\
g^B = Z_A^{-3/2}(g + \delta_g),\
v^B = Z_\Phi^{1/2}(v - \delta_v),
\end{equation}
where $Z_A = 1 + \delta_A$, $Z_\Phi = 1 + \delta_\Phi$ are the wave function
renormalization of the gauge boson and the link fields.

Let us first discuss the loop corrections to the mass of the massive gauge bosons $A^{(n)}_\mu$. Of course, there are no  reasons to expect that the loop corrections are finite, nevertheless,
since the set-up is renormalizable, all divergences can absorbed into  counterterms. 
From Eq.~(\ref{e.dct}) we find that the allowed counterterms corresponding to gauge boson masses are given by:
\begin{equation}
\cl_{ct}  =  \frac{1}{2}\, \delta_M g^2 v^2 \sum_{p=1}^N   (A_{\mu,p} - A_{\mu,p+1})^2  \, ,  
\end{equation}
where $\delta_M$ can be expressed  in terms of the wave function and gauge coupling renormalization as 
$\delta_M = 2 \delta_g/g + \delta_\Phi + \delta_A - 2 \delta_v/v $. 
Expressing the gauge fields in terms of the mass eigenstates these counterterms become: 
($N=2s+1$)%
\begin{equation}
\cl_{ct} = 
\frac{1}{2}\,  \delta_M \   \sum_{n=-s}^{s}  
 m_n^2 \, A_\mu^\bn  A_\mu^\bn 
\end{equation}
By adjusting  $\delta_M$ we can remove any divergence   proportional to $m_n^2$ that may appear in loop calculations of the gauge boson masses. The  finite part of $\delta_M$ depends on the regularization scheme, and therefore the renormalization of an overall scale of the gauge boson masses cannot be unambiguously calculated in deconstruction. On the other hand,  any loop corrections to the gauge boson masses that are not proportional to  $m_n^2$ (including a constant, $n$-independent, shift) are,  in  the deconstruction formalism, unambiguous predictions.

Consider now how  loop corrections to the mass of 4D massless scalar, {\it i.e.}, the zero mode of the fifth component of the gauge field $A_{5,\bz}$,  appear. 
To this end we need to analyze the possible form of the counterterms containing a  mass term for the
Goldstone boson $G_\bz$ and which descend from the counterterms involving the link fields $\Phi_p$. 
At the level of dimension $\leq 4$ operators and assuming  $N>4$, we have only the following `non-holomorphic' operators:
\begin{equation}
        \label{e.gmassdivergent}
\cl_{d} = \sum_{p=1}^N \delta_{d1} |\Phi_p|^2 + 
  \sum_{p,q=1}^N \delta_{d2} |\Phi_p|^2 |\Phi_{q}|^2   
\end{equation}
As a result of the translational invariance along the discrete lattice direction, $\delta_{d1}$ is independent of the lattice position $p$ while
$\delta_{d2}$ can only depend in the lattice distance $|p-q|$.
These operators can be induced with a divergent  coefficient. 
However, effectively, they do not introduce  any incalculable corrections to the mass of $G_\bz$. 
Indeed, once the link fields acquire a VEV, the Lagrangian~(\ref{e.gmassdivergent}) contains
both a mass term for the Goldstone boson $G_\bz$ and a tadpole
for the real part, $\Sigma_p$, of the link fields: 
\begin{equation}
        \label{e.nht}
\cl_{d} =    \delta_T  \left( \sfrac{2v}{N} \sum_p \Sigma_p +  G_\bz^2 \right) + \ldots 
\end{equation}
where $\delta_T$ is some function of the coefficients $\delta$'s in Eq.~(\ref{e.gmassdivergent}) --- note that the shift symmetry was essential to factorize the $\delta_T$ dependence
in~(\ref{e.nht}). Now adjusting the counterterms in order to remove the tadpoles automatically cancels  the mass term for $G_\bz$ as well. However the $G_\bz$ mass can be renormalized by gauge invariant `holomorphic' operators like, {\it e.g.}, 
 $\Phi_1 \Phi_2 \dots \Phi_N$. For $N>4$  the holomorphic operators  are  non-renormalizable and are induced at loop level with a finite, calculable coefficient. 
 We conclude that loop corrections to the $G_\bz$ mass are unambiguously calculable in deconstruction, once we fix the  counterterms such that   the $\Sigma_p$ tadpole term is vanishing. 

\section{Diagrammatic Computation}

\subsection{Mass corrections to $A_5$}

Let us start with computing the radiative correction to the mass of the
Goldstone boson that remains massless at tree-level.  Similar calculation, but in a non-renormalizable non-linear sigma model setup, was performed in ref. \cite{arcoge1}. As discussed in the previous section, in the renormalizable formalism  the first step is to calculate the diagrams that contribute to the tadpoles of the real part of link scalar fields, $\Sigma_p$, in order to determine the mass counterterm $\delta_T$, see~Eq.~(\ref{e.nht}). Then the mass correction of the physical Goldstone boson, $G_\bz$,  is obtained by 
calculating the two point function of this Goldstone mode and subtracting the contribution of $\delta_T$. 
The decomposition of the action in terms of the mass eigenstates leads to standard Feynman rules (see for instance~\cite{fgp}) which we can use to compute the two point function. After rather
long but trivial manipulations, we obtain:
\begin{equation}
        \label{e.d3}
\delta {m^2} =  - 4\, e_0^2 
\sum_{k=-(N-1)/2}^{(N-1)/2} 
\int  \frac{d^4 l_E}{(2 \pi)^4}\
\frac{ l_E^2 \cos ( 2 k \pi/N ) - m_k^2}{(l_E^2 + m_k^2)^2} 
\, .
\end{equation}
First we perform the momentum integration using dimensional regularization (we present
at the end of the Appendix a computation of the mass correction
where the summation over the KK modes is first performed).
Divergent terms cancel for $N>2$ \footnote{For $2< N \leq 4$ the behaviour of the two-point function is softer than expected from the discussion renormalizability because of the little-Higgs mechanism \cite{arcoge1}. However for $N>4$ the mass correction in deconstruction is calculable at any order of perturbation theory irrespectively of the little-Higgs arguments.}  and for the finite part we get: 
\begin{equation}
         \label{e.d4} 
\delta {m^2} = -\frac{4 e_0^2}{(4 \pi)^2}\, (2 e v)^2 
\left(  \vphantom{\frac{1}{2}}
- S_2(N) +  2 S_4(N) + 3 \Sigma_2 (N) - 4 \Sigma_4(N) 
\right) \, ,
\end{equation}
where the sums $S_{2m}$ and $\Sigma_{2m}$ are defined by ($N=2s+1$)
\begin{equation}
S_{2m} (N) = \sum_{k=-s}^{s} \sin^{2m}  \sfrac{k\pi}{N} 
\ \ \ \mathrm{and} \ \ \ 
\Sigma_{2m} (N) =
\sum_{k=-s}^{s} \sin^{2m} \sfrac{k\pi}{N} \, \log \sin^2  \sfrac{k\pi}{N}  .
\end{equation}
The sums $S_{2m}$ are trivially performed (see Appendix) and quite remarkably the sums $\Sigma_{2m} $ can also be performed analytically and they are expressed in terms of the  digamma function  $\Psi(z) \equiv { \Gamma'(z) \over \Gamma(z)}$ (see Appendix for details). So
the mass correction is finally written as:
\begin{equation}
         \label{e.d5} 
\delta {m^2} = -\frac{2 e_0^2}{(4 \pi)^2} \, (2 e v)^2 
\left( \vphantom{\frac{1}{2}}
\Psi (1+1/N)  - \Psi (1-2/N) +  \Psi (1-1/N) - \Psi (1+2/N)  
\right).
\end{equation}
By Taylor expanding the digamma function $\Psi$ around the unity, we easily obtain an $1/N$ expansion of the mass correction. In particular, using Eq.~(\ref{e.dge}), the leading terms in the correction are given by:
\begin{equation}
        \label{e.sums}
\delta m^2 = -  \frac{3 \, e_0^2}{4 \pi^2} \left( \frac{2 ev}{N} \right)^2 
\left ( \zeta(3) + {5 \zeta(5) \over N^2} + \dots \right) 
\ .
\end{equation}
Identifying  the parameters of the 5D and 4D theories as in Eq.(\ref{e.matching}) we can translate this result as:
\begin{equation}
\delta m^2 = -  \frac{3\, e_0^2}{4 \pi^4 R^2} 
\left ( \zeta(3) + {5 \zeta(5) \over (2 \pi \Lambda R)^2} + \dots \right) \, . 
\end{equation}
The first term  agrees with the mass correction~(\ref{e.dma5}) obtained by directly performing the  computations in the 5D theory~\cite{chmasc,geirqu,kupu}, while the second represents a correction due to a finite value of the 5D cutoff realized in the deconstruction setup. 

\subsection{Mass corrections to $A_\mu$}

Let us now turn to the more involved computation of the corrections to the gauge boson masses.
To this end, we need to evaluate the two point function of
the tree-level mass eigenstate gauge field $A_\mu^\bn$ which
we split into a transverse and a longitudinal part:
\begin{equation}
{\cal M}_n =  (p_\mu p_\nu - \eta_{\mu \nu}p^2) \Pi_1^n (p^2) 
+  \eta_{\mu\nu}\Pi_2^n (p^2).
\end{equation}
Then the shift of the mass at the $k^{\mathrm{th}}$ level  is given by: 
\begin{equation}
\delta {m_n^2} =  \Pi_2^n - m_n^2 \ \Pi_1^n. 
\end{equation}
After some algebra, the two form factors $\Pi_i^n$ are calculated to be ($N=2s+1$):
\begin{equation}
\Pi_1^n (p^2) =  8\, e_0^2 \sum_{k = -s}^{s} 
\int_0^1d x \ F_1^{n,k} (x)
\ \ \mathrm{and} \ \
\Pi_2^n (p^2) = - 4\,  e_0^2 \sum_{k = -s}^{s} 
 \int_0^1d x \ F_2^{n,k} (x)
\end{equation}
with
\begin{eqnarray}
F_1^{n,k} (x) & = & 
 \int \frac{d^d l_E}{(2 \pi)^d} 
\ \frac{x (1-x)}{(l_E^2 + x m_k^2 + (1-x) m_{n+k}^2 -x (1-x)p^2)^2},
\\
F_2^{n,k} (x) & = &
 \int \frac{d^d l_E}{(2 \pi)^d} 
\ \frac{(1 -2/d)l_E^2  +   m_k m_{n+k} \cos \sfrac{k\pi}{N}  - x (1-x) p^2}{
(l_E^2 + x m_k^2 + (1-x) m_{n+k}^2 -x (1-x)p^2)^2}.
\end{eqnarray}
In the previous integrals, $d=4$ is the dimension of the space-time and it will be promoted to $d=4-\epsilon$ in order to compute the integrals over the momenta using the usual recipes of dimensional regularization. The mass shift is then written as 
\begin{equation}
\delta m_n^2 = - \frac{2 e_0^2}{ (4 \pi)^2 }  
\left(
- \frac{1}{3} N \left( \frac{2}{\epsilon} -\gamma + \log (4\pi) \right) m_n^2  
+\sum_{k = -(N-1)/2}^{(N-1)/2} \int_0^1  d x \, f^{n,k}(x) \right)
\end{equation}
with
\begin{eqnarray}
f^{n,k}(x)= \left( m_{n+k}^2 +m_n^2 -m_k^2 +  2 x (m_k^2 - 2 m_n^2 - m_{n+k}^2) + 4 x^2 m_n^2 \right)
\hspace{1cm}
\nonumber\\
\log(  m_{n+k}^2  -x( p^2 - m_k^2 + m_{n+k}^2) + p^2 x^2 ).
\end{eqnarray}
Let us first note that the mass of the massless gauge boson does not get shifted ($\delta m_0^2=0$)
as a consequence of the unbroken $U(1)$ gauge symmetry. For the  massive gauge bosons 
($n \neq 0$),
the mass correction is divergent, but, as it should be according to our general analysis of the renormalization setup, the divergence is proportional to tree-level $m_n^2$ and so it  can be absorbed into counterterms. We keep only the finite  part  in the following formulae and  evaluate the mass correction on-shell, for $p^2 = m_n^2$. After integration over the Feynman parameter and
lengthy trigonometric manipulations and  after absorbing the finite terms
proportional to $m_n^2$ into the counterterms, we end up with the expression:
\begin{eqnarray}
&&
\delta m_n^2  =
 - \frac{2\,e_0^2 e^2 v^2}{3 \pi^2} 
\left( \vphantom{\frac{1}{2}}
S_2 (N) +  ( 1- \sfrac{3n\pi}{N} \sin \sfrac{2n\pi}{N} )\, S_4 (N)
- 2  \sfrac{n\pi}{N} \cot \sfrac{n\pi}{N} (1-4 \sin^2 \sfrac{n\pi}{N})\, S_6 (N)
\right)
\nonumber\\
&&
\hspace{2.5cm}
- \frac{\,e_0^2 e^2 v^2 }{\pi^2} 
\left( \vphantom{\frac{1}{2}}
\Sigma_2 (N) + 2\, \Sigma_4 (N) - 4\, \Sigma_6 (N)
\right) \, ,
\end{eqnarray}
where the sums are $S_{2m}(N)$ and $\Sigma_{2m}(N)$ have been defined previously, see   Eq.~(\ref{e.sums}). Using again the formulae from the Appendix to evaluate these sums, we obtain:
\begin{eqnarray}
        \label{e.amu}
&&
\delta m_n^2  =
- \frac{\,e_0^2 e^2 v^2 }{8\pi^2} 
\left( \vphantom{\frac{1}{2}}
 3\Psi (1+ \sfrac{1}{N})+3 \Psi (1- \sfrac{1}{N}) 
-4\Psi (1+ \sfrac{2}{N})-4\Psi (1- \sfrac{2}{N}) 
\right.
\nonumber\\
&&
\hspace{1cm}
\left. \vphantom{\frac{1}{2}}
+ \Psi (1+ \sfrac{3}{N}) +\Psi (1- \sfrac{3}{N})
\right) 
- \frac{\,e_0^2 e^2 v^2 }{24 \pi^2} 
\left( 
10 N - 9\, n \pi \cot \sfrac{n\pi}{N} -n \pi\, \frac{\cos \sfrac{3n\pi}{N}}{\sin \sfrac{n\pi}{N}}
\right).
\end{eqnarray}
The first term of the sum  does not depend on  the mass level $n$ and corresponds to the constant shift of the massive KK levels which, in the 5D setup, was found  in Ref.~\cite{chmasc}. The second term 
 does depend on $n$ and it appears here  because  deconstruction is a regularization that does not preserve 5D Lorentz invariance in UV. These terms however vanish when the continuum limit is taken. Indeed, using the expansion of the digamma function given in the Appendix, the leading terms in $1/N$ expansion of Eq.~(\ref{e.amu}) read
\begin{equation}
\delta m_n^2  = 
- \frac{e_0^2}{4 \pi^2} \left( \frac{ \vphantom{4} 2 e v}{N} \right)^2 
\left ( \zeta (3) - \frac{5\zeta (5)}{N^2} \right )
+ \frac{11\pi^2 \,e_0^2}{108} \left( \frac{ \vphantom{4} e v}{N} \right)^2 
\frac{n^4}{N}
+\ldots 
\end{equation}
which, in terms of 5D parameters, translates into
\begin{equation}
\delta m_n^2  = - \frac{e_0^2}{4\pi^4 R^2} 
\left( 
 \zeta (3) - \frac{5\, \zeta (5)}{( 2 \pi R\Lambda )^2}
 -\frac{11 \pi^3 \, n^4}{216 \, \Lambda R}  
+ \ldots
\right). 
\end{equation}

In the continuum limit $\Lambda \to \infty$ we recover
 the mass correction~(\ref{e.dmamu}) obtained by directly performing the
computations in the 5D theory~\cite{chmasc,geirqu,kupu}. But for a finite value of the cutoff  the correction  depends on the UV completion of the 5D theory. In particular, we can infer
that,  for a cutoff scale not much higher than the compactification scale, the prediction of the constant shift of the massive levels can be disturbed by UV physics, which may then play an important role for collider experiments.

\section{Operator analysis}

In Section 2 we signaled that operators responsible for the mass correction to the Goldstone boson $G_\bz$  are of the holomorphic structure $\Phi_1 \Phi_2 \dots \Phi_N$. From the 5D point of view such operators correspond to non-local Wilson lines winding around  the extra dimension. The renormalizable deconstruction setup offers thus a convenient setting to study loop induced non-local operators in a higher dimensional theory.  Indeed, the one-loop Coleman-Weinberg  potential for the gauge invariant phase 
$\phi \equiv {1 \over 2 i N} \log \left(\Phi_1 \Phi_2 \dots \Phi_N \over  \Phi_1^* \Phi_2^* \dots \Phi_N^* \right)$ can be expressed~\cite{arcoge1,hile} as: 
\begin{equation}
        \label{eq:CW}
V (\phi) = - \frac{e^4 v^4}{\pi^2} \sum_{k=-(N-1)/2}^{(N-1)/2}
\sin^4 \left ({k \pi \over N} + {\phi \over 2} \right)
\, \log  \sin^2 \left ({k \pi \over N} + {\phi \over 2} \right).
\end{equation}
Using the expressions for the sums $\Sigma_{2m}(N)$ introduced previously, we can easily find the Taylor expansion around $\phi=0$. In particular we can obtain this way the mass
of the Goldstone boson. Indeed at linear order, $\phi= G_\bz/(v \sqrt N)$, thus the quadratic term in this expansion of the effective potential~(\ref{eq:CW}) is directly related to the one loop mass of $G_\bz$.
We obtain
\begin{equation}
V (\phi) = cst - \frac{e^4 v^4}{\pi^2} 
\left( \vphantom{\frac{1}{2}}
6 \Sigma_2(N) - 8 \Sigma_4(N) + 7 S_2(N) - 8 S_4(N)
\right) 
\left( \frac{\phi}{2} \right)^2+ \ldots
\end{equation}
and using the formulae of the Appendix we end up for the mass of $G_\bz$ with the same expression (\ref{e.d5}) obtained
by a diagrammatic calculation.

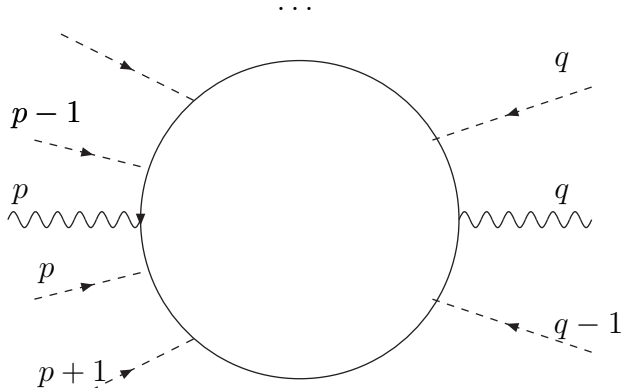
\begin{figure}
\begin{picture}(360,100)(0,0)
\Photon(90,50)(140,50){3}{6}
\DashArrowLine(110,120)(160,95){3}
\Text(105,90)[]{$p-1$}
\DashArrowLine(100,80)(140,70){3}
\Text(105,90)[]{$p-1$}
\DashArrowLine(100,20)(140,30){3}
\Text(105,30)[]{$p$}
\DashArrowLine(110,-20)(160,5){3}
\Text(115,-8)[]{$p+1$}
\ArrowArc(200,50)(60,0,360)
\Photon(260,50)(310,50){3}{6}
\DashArrowLine(310,100)(250,80){3}
\Text(300,110)[]{$q$}
\DashArrowLine(310,0)(250,20){3}
\Text(310,10)[]{$q-1$}
\Text(200,130)[]{$\dots$}
\Text(300,60)[]{$q$}
\Text(95,60)[]{$p$}
\vspace{1cm}
\end{picture}
\caption{One-loop diagram contributing to the mass shift of the KK gauge bosons. In the 5D language, an effective non-local operator involving derivative of Wilson lines is generated.}
\label{fig.wilson1}
\end{figure}

Quite analogously, the constant shift of the heavy gauge boson mass levels can be ascribed to holomorphic operators that are interpreted as non-local from the 5D point of view. For instance, the diagram in Fig.~\ref{fig.wilson1} induces an  operator of the form:
\begin{equation}
        \label{e.wilson1}
\cl \sim ( A_{\mu,p} \Phi_p \dots \Phi_{q-1}  A_{\mu,q} \Phi_q \dots \Phi_{p-1})
\end{equation}
This operator is invariant only under  global transformations of the product group and so it must be a part of some locally invariant operator. Let us define the `Wilson-line' operators, 
$W(p,q) \equiv \Phi_p \dots \Phi_{q-1} $,
and their covariant derivatives, 
$D_\mu W(p,q) \equiv  \pa_\mu W(p,q)  + i e A_{\mu,p} W(p,q) - i e W(p,q) A_{\mu,q}$. 
Then  locally (and shift symmetry) invariant operator which contains that of Eq.~(\ref{e.wilson1}) is given by:
\begin{equation}
        \label{e.wilson2}
\cl \sim \sum_{p,q=1 p \neq q}^{N} D_\mu W(p,q) D_\mu W(q,p) + \hc 
 \end{equation}
when the links get VEVs such operators yield mass terms for the gauge bosons of the form 
\begin{equation}
        \label{e.mass.dqed}
\cl \sim v^2 \sum_{p,q=1}^{N} (A_{\mu,p} -A_{\mu,q})^2 
\end{equation}
Inserting the mode decomposition for $A_{\mu,p}$  we get precisely the  constant shift of the massive KK modes
\begin{equation}
        \label{e.mass6}
\cl \sim v^2 \sum_{n \neq 0} A_\mu^\bn A_\mu^\bn
\end{equation}
The 5D non-local operator that corresponds to the deconstructed operator of (\ref{e.wilson2})
involves covariant derivatives of the Wilson lines
\begin{equation} 
        \label{e.wilsoncd5d}
S \sim \frac{1}{R^4} 
\int d^4x \, \int_0^{2 \pi R}dy_1 dy_2\,  
D^\mu (e^{i \int_{y_1}^{y_2} d \t y g_5 A_5})  
D_\mu (e^{i \int_{y_2}^{y_1} d \t y g_5 A_5})   \, ,    
\end{equation}
This operator yields  a constant shift of the massive KK gauge bosons of the 
form~(\ref{e.dmamu}).  However to be able to determine the exact value of the mass shift, one should compute non only the coefficient of the operator~(\ref{e.wilson1}) but also the coefficients
of infinite number of other holomorphic operators, like for instance
\begin{equation}
\cl \sim ( A_{\mu,p} \Phi_p^k \dots \Phi_{q-1}^k  A_{\mu,q} \Phi_q^k \dots \Phi_{p-1}^k),
\end{equation}
and non-holomorphic operators like
\begin{equation}
\cl \sim ( A_{\mu,p} \Phi_p \dots \Phi_{q-1}  A_{\mu,q} \Phi_q \dots 
\Phi_{r-1} |\Phi_r|^{2k}\Phi_{r} \dots \Phi_{p-1}).
\end{equation}
Whether it exists an appropriate choice of variable, like in~(\ref{eq:CW}), that allows
to sum all those  operators is an open question that deserves further scrutiny.

In any case the 4D analysis leads to an identification of  non-local operators that are responsible for the mass shift of both $A_\mu$ and $A_5$ in five-dimensional gauge theories .

%

\section{Conclusions}

In this paper we calculated one-loop corrections to the Kaluza-Klein gauge boson excitations in the deconstructed version of the 5D QED. The deconstructed set-up, being a renormalizable UV completion of the 5D theory, is a useful framework for studying quantum corrections. Moreover, it enables to control the cut-off dependence of  5D theories and study a possible influence of UV physics on IR observables.   
 Our results are consistent with those obtained in refs \cite{chmasc,geirqu,kupu} by direct computations in the 5D theory. We  calculate the $\Lambda$-dependent non-leading corrections and point out  that sensitivity of the 5D theory to UV physics may be phenomenologically relevant. We also discuss the structure of  operators that are relevant for the quantum corrections to the gauge boson masses in 5D and in deconstruction.

\renewcommand{\theequation}{A.\arabic{equation}}
\setcounter{equation}{0}

\section*{Appendix: Reference Formulae}
\label{app}

In this appendix we present formulae for various sums appearing in diagrammatic computations  and we collect various properties of the digamma function.

The  sums, $S_{2m}$, involving  even powers of sines can  be computed using a Chebychev decomposition of $\sin^{2m} \theta$:
\begin{equation}
S_{2m} (2s+1)
=
\sum_{k=-s}^s \sin^{2m} \left( \sfrac{k\pi}{2s+1} \right)
= \sfrac{(2m-1)!!}{(2m)!!} (2s+1) \, .
\end{equation}
The sum $\Sigma_{2m}$ is defined as
\begin{equation}
\Sigma_{2m} (2s+1) =
\sum_{k=-s}^{s} \sin^{2m} \left( \sfrac{k\pi}{2s+1} \right) \log \sin^2 \left( \sfrac{k\pi}{2s+1} \right) \, , 
\end{equation}
and it can be performed analytically by the use of the Gauss' theorem about the digamma function. For $0<p<2s+1$ we have:
\begin{equation}
\Psi \left( \sfrac{p}{2s+1} \right)
= -\gamma - \log (4s+2) -\sfrac{\pi}{2} \cot \left( \sfrac{p\, \pi}{2s+1} \right)
 + \sum_{k=1}^{s} \cos \left( \sfrac{2pk \pi}{2s+1}  \right) 
 \log  \sin^2 \left( \sfrac{k\pi}{2s+1} \right) 
\end{equation}
Here $\gamma \sim .577\ldots$ is the Euler--Mascheroni constant and  $\Psi (z)$ stands for the digamma function, which is defined as  the logarithmic derivative of the Euler gamma function, $\Gamma (z)$:
\begin{equation}
\Psi (z) = \Gamma^\prime (z) / \Gamma (z). 
\end{equation}
From  the Gauss' digamma theorem one can derive the general expressions ($0<2m<N$):
\begin{eqnarray}
& \displaystyle
\Sigma_{0} (N)  =  \log N - (N-1) \log 2,
\\
& \displaystyle
\Sigma_{2m} (N)  =  
 \frac{1}{2^{2m-1}} 
\left(
-\binomial{2m}{m} (\gamma + N \log 2)
+
\sum_{k=1}^m (-1)^k \binomial{2m}{m-k} \left(2 \Psi (\sfrac{k}{N}) 
+ \pi \cot (\sfrac{k\pi}{N}) \right)
\right) .
\end{eqnarray}
In particular, using  the following relations about the digamma function
\begin{eqnarray}
        \label{e.dgi}
& \displaystyle
\Psi(z) = \Psi(1-z) - \pi \cot(\pi z),
\\
& \displaystyle
\Psi(1+z) = \Psi(z) + \frac{1}{z},
\\
& \displaystyle
\Psi(1) = - \gamma,
\end{eqnarray}
one obtains:
\begin{eqnarray}
\Sigma_2(N) &=& -{1 \over 2}\Psi(1+{1 \over N})-{1\over 2}\Psi(1-{1 \over N})  - (N \log 2+\gamma) + {N \over 2};
\\ 
\Sigma_4(N) &=& -{1 \over 2}\Psi(1+{1 \over N})-{1 \over 2}\Psi(1-{1 \over N})  +{1 \over 8}\Psi(1+{2 \over N}) + {1 \over 8}\Psi(1-{2 \over N});
\nn
 &&
\hspace{.5cm}
- {3 \over 4}(N \log 2+\gamma) + {7 N \over 16}
\\ 
\Sigma_6(N) &=&
 -{15 \over 32}\Psi(1+{1 \over N})-{15\over 32}\Psi(1-{1 \over N}) 
 +{3 \over 16}\Psi(1+{2 \over N})+{3 \over 16}\Psi(1-{2 \over N})
\nn &&
\hspace{.5cm}
 -{1 \over 32}\Psi(1+{3 \over N})-{1\over 32}\Psi(1-{3 \over N}) 
- {5 \over 8}(N \log 2+\gamma) + {37 N \over 96}.
\end{eqnarray}
In order to find the $1/N$ expansion of these results we introduce 
the $n^\mathrm{th}$ polygamma function, $\Psi^{(n)} (z)$, which is defined as the $(n-1)^\mathrm{th}$ derivative of the $\Psi (z)$ function. From the series representation of the $\Gamma$ function, the polygamma function
can be related to the Hurwitz $\zeta$ function defined by 
$\zeta(s,a)=\sum_{k=0}^{\prime\, \infty} (k+a)^{-s}$ (the prime meaning that the possible value of $k$ such that $k+a=0$ is omitted in the sum)
\begin{equation}
\Psi^{(n)} (z) = (-1)^{n+1}  n!\,  \zeta (n+1,z). 
\end{equation}
In particular, we get that 
\begin{eqnarray}
        \label{eq:zeta3}       
&&
\Psi^{(2)} (1) = - 2!\, \zeta (3) \, ,
\\
&&
\Psi^{(4)} (1) = - 4! \, \zeta (5) \, ,
\end{eqnarray}
where $\zeta (s)=\sum_{k=1}^{\infty} k^{-s}$ is the usual Riemann $\zeta$ function. We thus find:
\begin{equation}
        \label{e.dge}
\frac{1}{2} 
\left(
\Psi(1+ \frac{a}{N}) + \Psi(1-\frac{a}{N})
\right) 
 = 
 - \gamma - \frac{\zeta(3) a^2}{N^2} - \frac{\zeta(5) a^4}{N^4} + \dots 
\end{equation}

Let us finally mention that we can alternatively compute
the mass correction (\ref{e.d5}) to $A_5$ by first performing 
the summation over the KK mode in Eq.~(\ref{e.d3}) and then performing the
momentum integration.  To this end, the following sum is needed
\begin{equation}
        \label{eq:trigsum}
\sum_{k=-(N-1)/2}^{(N-1)/2}
\frac{1}{\sinh^2 x + \sin^2 k\pi/N} =
\frac{2N \mathrm{cotanh} Nx}{\sinh 2x}.
\end{equation}
This relation can be proved by a pole decomposition of the right hand side.
And the resulting momentum integration reduces to
\begin{equation} 
 \int_0^\infty dx\ 
\frac{\sinh^3 (x/2)\cosh (x/2)}{\sinh^2 (N x /2)}
=
\frac{1}{2N^2} 
\left( \vphantom{\frac{1}{2} } 
\Psi (\sfrac{N+1}{N})  - \Psi (\sfrac{N-2}{N}) +  \Psi (\sfrac{N-1}{N}) - \Psi (\sfrac{N+2}{N}) 
\right).
\end{equation}
%

\section*{Acknowledgments}

We are grateful to Luigi Pilo for collaboration at the early stage of this project and numerous discussions later on.  We also thank Martin Puchwein and Kirone Mallick for turning our attention to the Gauss' digamma theorem. AF thanks Yasunori Nomura for interesting discussions.  
The three of us  were funded in part by the RTN European Program HPRN-CT-2000-00148.   A.F. was partially supported Polish KBN grant 2 P03B 070 23 for years 2002--2004.  C.G. was also supported in part by  the ACI Jeunes Chercheurs 2068. S.P. was partially supported Polish KBN grant 2 P03B 129 24   for years 2003--2004. S.P. thanks Institute of Physics in Bonn for its hospitality.
His visit to Bonn was possible thanks to the research award of the
Humboldt Foundation.


\end{document}